\documentclass[journal]{IEEEtran}

\usepackage{caption2}
\usepackage{graphicx}
\usepackage{booktabs}  
\usepackage{amsmath}
\usepackage{mathrsfs}%花体字母
\usepackage{cite}
\usepackage{verbatim}
\usepackage{amssymb}
\usepackage{enumerate}
\usepackage{algorithmicx,algorithm}
\usepackage[noend]{algpseudocode}
\usepackage{multirow}
\usepackage{color}
\usepackage{fancyhdr}

\begin{document}

\title{A Novel Fuzzy Search Approach over Encrypted Data with Improved Accuracy and Efficiency}

\author{Jinkun Cao,
        Jinhao Zhu,
        Liwei Lin,
        Zhengui Xue,
        Ruhui Ma*\thanks{* = Corresponding author},
        Haibing Guan
\\ \textbf{Shanghai Jiao Tong University, Shanghai, China}
\\ \emph{\{caojinkun, zhujinhao, llw02\_02 , zhenguixue, ruhuima , hbguan \}@sjtu.edu.cn}}

\markboth{}%
{Shell \MakeLowercase{\textit{et al.}}: Bare Demo of IEEEtran.cls for IEEE Journals}

\maketitle

\begin{abstract}
As cloud computing becomes prevalent in recent years, more and more enterprises and individuals outsource their data to cloud servers. To avoid privacy leaks, outsourced data usually is encrypted before being sent to cloud servers, which disables traditional search schemes for plain text. To meet both end of security and searchability, search-supported encryption is proposed. However, many previous schemes suffer severe vulnerability when typos and semantic diversity exist in query requests. To overcome such flaw, higher error-tolerance is always expected for search-supported encryption design, sometimes defined as 'fuzzy search'. In this paper, we propose a new scheme of multi-keyword fuzzy search over encrypted and outsourced data. Our approach introduces a new mechanism to map a natural language expression into a word-vector space. Compared with previous approaches, our design shows higher robustness when multiple kinds of typos are involved. Besides, our approach is enhanced with novel data structures to improve search efficiency. These two innovations can work well for both accuracy and efficiency. Moreover, these designs will not hurt the fundamental security. Experiments on a real-world dataset demonstrate the effectiveness of our proposed approach, which outperforms currently popular approaches focusing on similar tasks.
\end{abstract}


\IEEEpeerreviewmaketitle

\section{Introduction}

% 第一段，说明为什么现在很多的数据要在cloud上面处理
\IEEEPARstart{I}{n} the age of ``Big Data'' and mobile Internet, the volume of data produced and processed through Internet expands fiercely. It brings a higher demand for storage and computation capacity. Since personal devices are incapable of handling it on many occasions, cloud service is endowed with more importance\cite{dinh2013survey,qureshi2011mobile,namboodiri2012cloud}. Meanwhile, cloud storage and computing bring new concern on privacy protection\cite{sun2014data,chen2012data}.

% 第二段：介绍现在流行的三方"加密-搜索-返回"模型和searchable encryption的背景
To protect data privacy, outsourced data is usually encrypted in advance by data owners, after which the data query and search are performed. However, conventional search methods cannot be implemented in the ciphertext. Thus, search-supported encryption is proposed\cite{encryptbeforeoutsource}, with which relevance degree among encrypted text can be measured and thus search over encrypted data becomes possible. Furthermore, considering ambiguity, typos, grammar variance, and semantic variety, bias is common for text matching. Fuzzy search is thus proposed\cite{earlyfuzzy, target, firstfuzzy} to achieve more robust search performance with these noises involved (e.g. misspelling detection feature of search engines). Facing the same demand, fuzzy search is also developed on encrypted data. As usual scenes of data search, users care most about the accuracy and efficiency of searching. With a metric, there should be a ground truth rank between stored files and input query by their similarity. Correspondingly, a higher accuracy for search over encrypted data asks that the set of returned files should share a larger overlap with the ground truth most similar files. Efficiency is reflected by the time latency during search and obviously depends on the search algorithm performance. In this paper, we propose a novel multi-keyword fuzzy search design over encrypted cloud data. We promote performance on both accuracy and time efficiency with different innovations.

We conclude the pipeline of searchable encryption approaches into three steps as follows:

\begin{enumerate}
	\item Represent: Keywords are extracted from outsourced files or received queries, and transfer into word-vectors, a combination of which builds the final representation of files or queries.
	\item Encrypt and index: Files and queries are both encrypted to enhance security. They are suggested to be encrypted in heterogeneous ways. The encryption algorithm and key are usually provided by data owners. With some data structure, encrypted files are organized and stored for indexing.
	\item Search: In practice, data users send queries and data holders perform some search algorithms on the query and stored encrypted data. Search consists of the calculation of relevance score and ranking by the score. The data user usually only asks for the top-k most relevant files with the query instead of all relevant files.
\end{enumerate}

% 第四段：介绍fuzzy search的技术细节
The first step is critical when designing a fuzzy search mechanism. By representing files in a certain structure, keyword information within files is mapped into a uniform representation space. Similar files, mostly containing a close composition of keywords, are expected to be mapped close in this space. The representation should be error-tolerant for common language bias, such as typos and synonyms, to return users' desired results. On the other hand, the search index, encryption techniques, and search algorithms should cooperate well to conduct fast and correct retrieval. 

% 第五段：介绍现在已有的fuzzy search方法的不足
Even though, promising search accuracy is achieved by current fuzzy search schemes on some occasions, it may fail in many others. Typos, grammatical bias, semantic diversity often bring troubles. Non-perfect representation schemes even create new troubles. For example, anagrams (different words consisting of the same set of letters) will puzzle some schemes\cite{target} to map different keywords into the same location in the representation space. To overcome similar flaws and promote search accuracy, we propose a novel text presentation scheme that generating keyword-vectors based on designed 'order-preserved uni-gram' (OPU). OPU outperforms popular 'uni-gram'\cite{target} of 'n-gram'\cite{firstfuzzy} in term of accuracy in many cases. The keyword pair 'silent'/'listen' (anagram) or 'keep'/'keap'(typo) has no chance to confuse our proposed mechanism as it can do in some other cases.

% 第六段：介绍在数据结构、搜索算法方面做的创新，提高效率
Because efficiency is another major concern during data search, accelerating search on without too much harm to accuracy is expected. We also propose an improvement on this aspect by renewing the design of data structure and search algorithm. Precisely, we propose an improved data organization scheme and design search algorithm based on it. A novel data clustering method is designed to gather similar files within a cluster. These clusters are some continuous areas in the aforementioned representation space. Furthermore, an index tree is built in a hierarchical manner by organizing those file clusters. Namely, we design a hierarchical index tree (HIT) for data organization. Compared with the previous designs\cite{treesearch1, treesearch2, treetopdown, minihash}, this design is expected to achieve better time efficiency with little harm to accuracy. Moreover, it is flexible enough to adapt to different cases with less hand-adjusted parameters. Moreover, such tree-based data organization brings extra convenience to do verification\cite{verifiable1, verifiable2} after data retrieval to ensure the freshness, correctness, and completeness of returned data.

% 下面这一个部分挪到related work关于search算法使用的数据结构的介绍部分。

At last and most importantly, security and privacy should be guaranteed in our proposed architecture, which is expected to be ensured under different popular threat models\cite{threatmodel1}. Therefore, focusing on the problem of fuzzy multi-keyword search over encrypted data, we summarize our contributions proposed in this paper in term of two aspects:

\begin{itemize}
	\item We improve accuracy under many cases by designing a novel file representation scheme named 'order-preserved uni-gram' (OPU). It maps similar text to be close in representation space with kinds of noise involved. 
	\item We improve the search efficiency by designing a new data structure for data organization and corresponding search algorithms. Hierarchical index tree (HIT) is adopted in our scheme to improve time efficiency during query with slight harm to accuracy. To organize outsourced data, an improved dynamic clustering algorithm is proposed which needs less pre-set parameters and thus more flexible.
\end{itemize}

These two innovations are involved in different stages of search on encrypted data and contribute to the final proposal differently. Of course, our proposal faces the trade-off between time efficiency and search accuracy: OPU brings slightly more computation overhead during the file processing and indexing stage than the word parsing based on naive ``uni-gram''; construction of HIT brings slight negative effect on the search accuracy as well. However, both of the degradations are trivial enough that the overall design benefits both accuracy and time efficiency. We design various experiment to compare our proposed approach; with state-of-the-art schemes\cite{target} on real-world linguistic dataset\cite{20newsgroup} and the result proves the effectiveness of our approach.

\section{Related Work}
\subsection{Searchable Encryption}
Curtmola et al. \cite{securitydefine} proposed a security definition on searchable encryption which is followed by most popular mechanisms. Song et al.\cite{song2000} proposed the first practical searchable encryption mechanism. Previous work focused on improving search accuracy and efficiency without harm to a necessary security guarantee. Wang et al.\cite{Wang2010Secure} made an important improvement by introducing novel file indexing techniques. Cao et al.\cite{cao2014} proposed a novel encrypted search scheme supporting multi-keyword matching by coordinate matching. Focusing on reaching similar demand, conjunctive keyword search\cite{conjuntive2014}, vector-space-based search models\cite{vectorspace} and many other works were proposed. Many aforementioned studies focused on finding a higher efficient file representation design. To the end of encryption, ``Secure kNN'' algorithm\cite{secureknn2} is adopted in most currently popular approaches\cite{secureknn, secureknn2, target, Wang2010Secure}.
	
\subsection{Quicker search}
It is costly to traverse the whole content of all files to calculate their similarity with the query. To tackle the problem of quicker comparison on large-scale file storage, file indexing technique usually is adopted. As the first step, a file is represented by extracted keywords from it, which compresses the indexing volume to a great extent. Then it is common to use some data structures to organize these compressed file indices, on which previous encrypted search schemes propose many innovations. Linear search by a single thread on indices always provides the lower bound of time efficiency because all files are traversed. As an improvement, some studies\cite{parallel1, parallel2} try to improve latency through parallel computation and multi-task distribution. Some others focus on the improvement of the data structure for file indexing: hashing\cite{minihash} and tree-like data structure\cite{treesearch1, treesearch2, treetopdown} are widely researched for this topic. Recently, Chen et al.\cite{HCandVeri} introduced Hierarchical Clustering mechanism into encryption data organization and search algorithm. It is unavoidable in most cases that efficiency improvement from novel data structure brings harm to search accuracy because some files are skipped during the search to save time. To improve time efficiency without too much harm to accuracy is also expected in this area.

\subsection{Fuzzy Search}
Language bias such as misspellings and multiple-semantic expression is common and ought to be recognized to improve search accuracy. However, it is hard to distinguish many confusing pairs of words to be typos or different words. For example, how an automated system recognizes that received ``catts eat mice'' is a typo of ``cats eat mice'' and search for stored data relevant to the latter and wanted expression? Furthermore, encryption makes it even much more difficult because a slight difference in original natural language expression could be heavily zoomed out after encryption. When calculating the similarity of different text, these recognized bias declines the search accuracy. To enhance the robustness for search, it comes to the issue of ``fuzzy search'' topic. Li et al.\cite{earlyfuzzy} first formalized and enabled the fuzzy search over encrypted data while maintaining the security guarantee with the help of a pre-defined fuzzy set. Under similar pre-defined dataset, Fu et al.\cite{personalized} improved search accuracy when synonyms or antonyms exits in text. Without the assistance of extra information, fuzzy search usually is designed based on text representation and indexing schemes with high language-bias-tolerance. Wang et al.\cite{firstfuzzy} relieved the effect of typos or grammatical diversity by transforming keywords into \emph{'bi-grams'}. Furthermore, Fu et al. transform keywords into \emph{'uni-grams'} in \cite{target} to achieve better performance. But word parsing method based on \emph{'uni-grams'} in \cite{target} still have many defects such as not enough robust when anagrams, special character or some other kinds of language bias are involved in text materials.

\section{Preliminaries}
\subsection{System Model}
As popular designs\cite{earlyfuzzy,cao2014,target,firstfuzzy}, the system model our scheme faces consists of three components: data owner, cloud server and data user. Data owners encrypt files before outsourcing them and build a data structure to index these outsourced files. The file indices are also encrypted. Outsourced data is only exposed to certified data users and trusted remote servers. Certified data users encrypt their queries and send them to remote servers. Search is operated on remote servers to leverage its computation and storage capacity. Relevance scores between queries and the stored data are computed with some algorithms. Remote servers then return $top-k$ most relevant files to data users. Data users would decipher these files with keys from the data owner.

\begin{figure}[htp]
	\centering
	\includegraphics[width=3.5in]{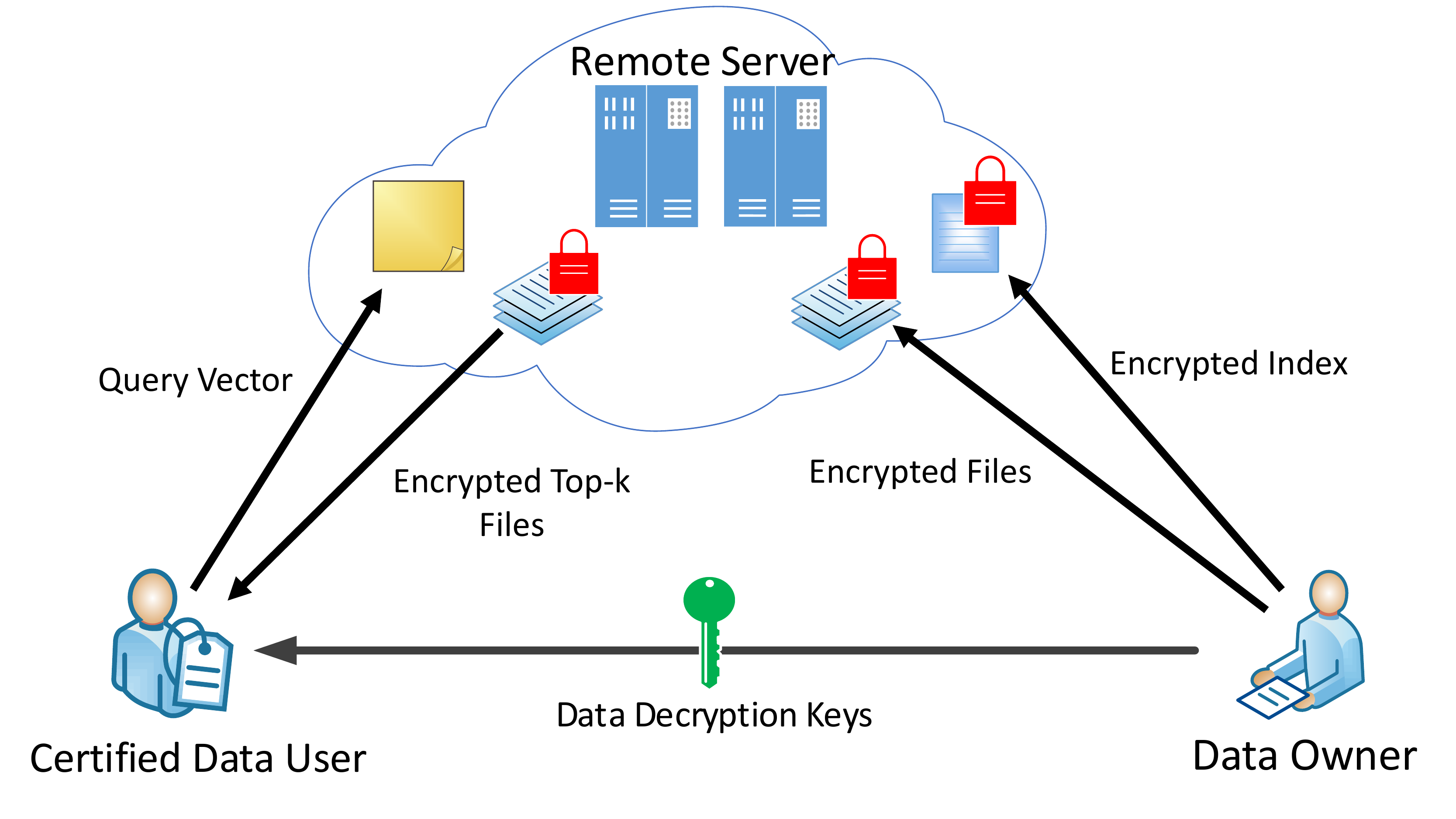}
	\caption{System Model with three parties}
	\label{systemmodel}
\end{figure}

\subsection{Locality-Sensitive Hashing (LSH)}
As a subclass of hash functions, locality-sensitive hashing functions have a significant feature. More similar items, which can be determined with some distance metric $d$ (e.g. Euclidean distance), are more likely to be hashed into the same group. A LSH family $\mathcal{H}=\{h_1,h_2,...,h_l\}$ is defined as $(r_1,r_2,p_1,p_2)-sensitive$ if for each $h_q\in\mathcal{H} (1{\le}q{\le}l)$, two arbitrarily chosen points $u$,$v$ satisfy:
\begin{subequations}
\begin{align}
	if \ d(u,v)\leq r_1: Pr[h_q(u)=h_q(v)]\geq p_1 \\
	if \ d(u,v)\geq r_2: Pr[h_q(u)=h_q(v)]\leq p_2 
\end{align}
\end{subequations}
where $d(u,v)$ is the distance between these two points to represent their similarity. $p-stable$ LSH \cite{pstablelsh} is a specific kind of LSH methods based on a $p-stable$ function family, which can be formulated as:
\begin{equation}
	h_{a,b}(\overrightarrow{v})=\lfloor \frac{\overrightarrow{R}\cdot\overrightarrow{v}+b}{a} \rfloor
\end{equation}
where $\overrightarrow{R}$, $\overrightarrow{v}$ are two vectors and $a$ and $b\in{[0, a]}$ are two real numbers. $\overrightarrow{R}$, $a$ and $b$ are parameters and $\overrightarrow{v}$ is the variable to be hashed.

\subsection{Bloom Filter}
Bloom Filter\cite{bloomfilter} is a special data structure widely adopted to map a high-dimensional point into a space of lower dimensions. A D-dimensional point is transformed into a one-hot m-bit vector with a hash function. With a set of $l$ independent hash functions $\mathcal{H}=\{h_1,h_2,...,h_l\}$, the point can be thus transformed into a m-bit vector with at most $l$ nonzero bits. With the features of $LSH$, more similar keywords are expected to be mapped into the same position with higher probability by the same $LSH$ function. Thus the finally generalized vectors are more likely to be similar or even the same. For a given set of points $\mathcal{P}=\{p_1,p_2,...,p_n\}$, the $l$ independent LSH functions encode each $p_i(1 \le i \le n)$ into at most different $l$ bits of a m-bit vector $BF_{\mathcal{P}}$. This vector is thus called 'Bloom Filter'. To judge whether a point $p\in \mathcal{P}$, we simply generalize its corresponding bloom filter the same set of hash functions and test whether there is the same bloom filter for $p_i$ found. As proved in\cite{bloomfilter}, the chance to give a false positive through this method is approximately $(1-e^{-\frac{l\times n}{m}})^{l}$. The minimal rate of false positive is $(\frac{1}{2})^{l}$, which is achieved when $n\times l=m\times \ln{2}$. Thus a better expected false positive is available with bigger $l$. But to set $l$ small can keep produced Bloom Filter sparse, which is helpful to increase the accuracy of our scheme. This raises an important trade-off for application of bloom filters.

\subsection{Hierarchical Clustering Tree}
Clustering algorithms are adopted to divide items into different clusters through comparing their similarity, namely items are divided into the same cluster if they are adjacent enough in the vector space. Popular hierarchical clustering techniques, such as k-means\cite{kmeans}, DBSCAN\cite{dbscan} and GMM\cite{gmm}, are widely adopted in data mining. Hierarchical clustering in\cite{HCandVeri} sets a maximum number of elements in a cluster and then begins with random sample points to divide adjacent items into one cluster. For data organization in search cases, the element to be clustered can be a file or a query or the center point of a sub-cluster, all of which are first mapped to a uniform space and represented by a point in the space. In hierarchical clustering, the clustering algorithm is performed recursively on original elements and the points representing sub-clusters. Finally, we could generalize a hierarchical clustering tree. Specifically, each outsourced file is a leaf node of the hierarchical clustering tree built through. To search on the tree to find some nodes, the general time complexity is simply $\mathcal{O}(log(n))$ instead of $\mathcal{O}(n)$ for linear search.

\begin{figure} 
	\centering
	\includegraphics[width=3.3in]{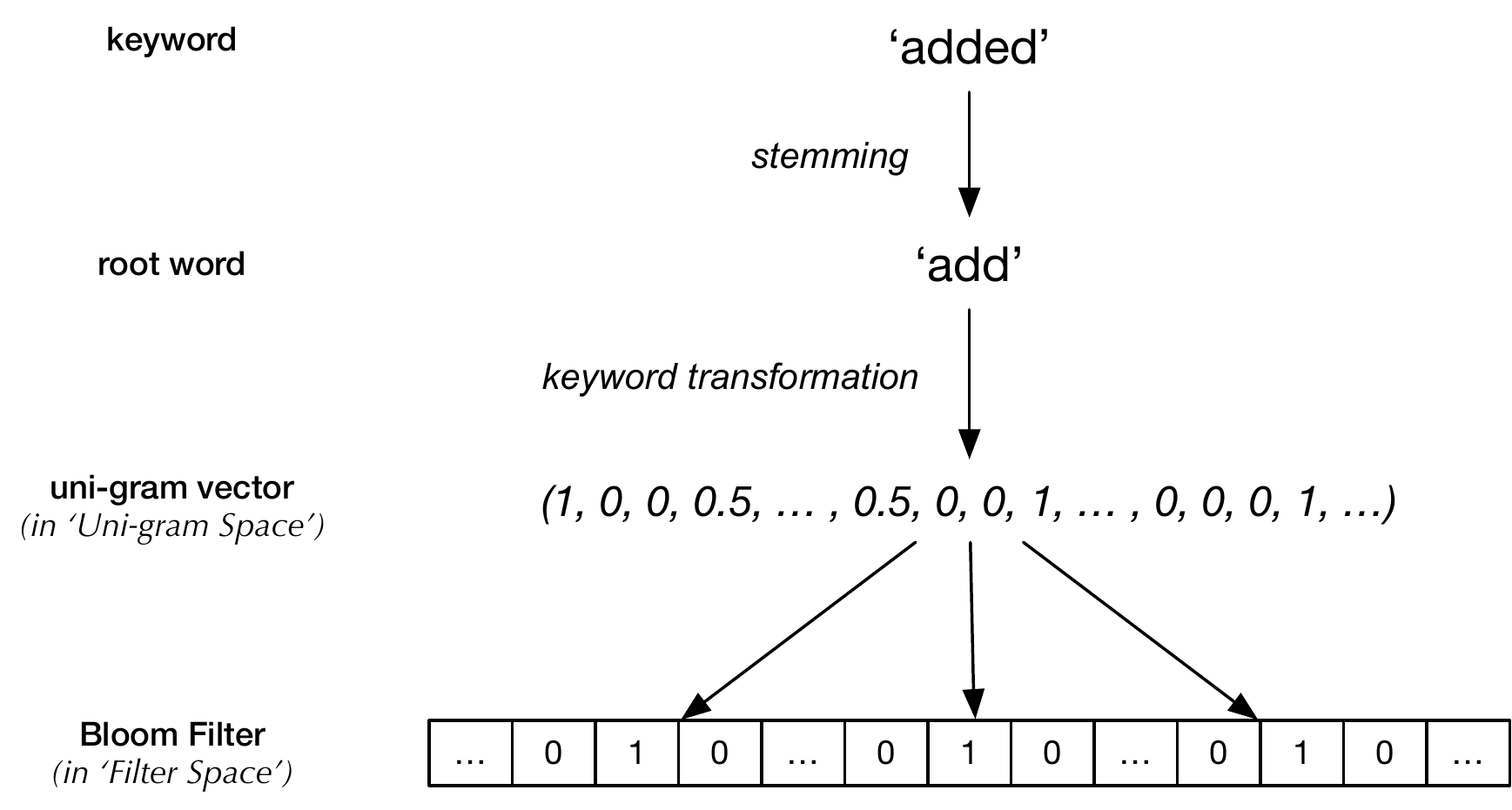}
	\caption{Transformation from keyword to Bloom Filter}
	\label{opug}
\end{figure}

\section{Proposed Algorithms}
In this section, we introduce our proposed algorithms in detail. Our main innovations include a novel keyword transformation scheme and a specially designed data structure for indexing. The former is named 'order-preserved uni-gram'(OPU) to show its difference with traditional 'uni-gram'\cite{target} or 'n-gram'\cite{Wang2010Secure}. The designed data structure is called ``Hierarchical Index Tree'' (HIT). With the help of OPU, more information of the original natural language keywords can be reserved without privacy leaks, providing help for more accurate search. On the other hand, we propose an adaptive clustering algorithm to build HIT and a corresponding search algorithm to better balance the trade-off between search efficiency and accuracy.

\subsection{``Order-Preserved Uni-gram'' (OPU)}
Misspelling occurs in three cases: \emph{'misspelling of letters', 'order of letters reversed' and 'addition or missing of letters'}. To judge the similarity of two keywords, the popular mechanism usually splits keywords into some ``pieces''. The granularity of which keywords are decomposed influence the information it can memorize and the collision chance that different keywords are transformed into similar or even the same set of pieces. Wang et al.\cite{firstfuzzy} adopted the \emph{'n-gram'} method, which would split the word \emph{'task'} into the set $\{ta, as, sk\}$ when $n=2$. This method achieves a good result for the first case but fails for another two cases. Based on this scheme, Fu et al.\cite{target} proposed \emph{'uni-gram'} method, under which, for instance, the uni-gram set of keyword \emph{'scheme'} is $\{s_1,c_1,h_1,e_1,m_1,e_2\}$. This scheme performs better for the other two cases. However, due to the lack of information on the order of letters, such a scheme is incapable of recognizing \emph{'anagram'}, such as 'devil' and 'lived'.

To bring better fuzzy search, we propose a new method to transform a keyword into an 'order-preserved uni-gram' vector, which would show advantages compared with previous work. While keywords are of different length, the output OPU vectors are expected to be equally long. The construction of an OPU vector can be divided into three steps, decompose, encode and infect, introduced as follows:

\subsubsection{Decompose}
Given a set of keywords extracted from some text materials, we perform this operation on each of them respectively. First, all keywords should be stemmed\cite{stemming} to eliminate the grammatical and other linguistic variations. Then all keywords would be dismembered into single letters. But different from the unordered set of letters for 'uni-gram', we also record the position of each letter in the original keyword for further use.

\subsubsection{Encode}

 To relieve computation and storage overheads, at most first $u$ letters in a keyword can be fully represented after transformation. The corresponding length of output OPU vector, $V_{KW}$, should be $(u\times26+30)$, all bits of which are binary. Specifically, the vector consists of $u$ ``letter blocks''(LB), each of which has 26 bits, and one 'digit and symbol block'(DSB), whose length is 30 bits. Letters between 'a' to 'z' are mapped to $1^{st}-26^{th}$ bit in each letter block. The $i^{th}$ letter in $KW$ corresponds to $i^{th}$ letter block in $V_{KW}$. 
 
 For example, if the target keyword is ``add'', only $V_{KW}[1]$, $V_{KW}[26+4]$ and $V_{KW}[26 \times 2 + 4]$ are set to 1 with all other bits remaining 0 (because ``a'' and ``d'' are respectively the first and the fourth letter in alphabet). The last $30$ bits in $V_{KW}$ indicate whether 10 digits ('0'-'9') and 20 widely used symbols appear in $KW$. In practice, if a keyword is too long to fit the preset length of an OPU vector, it will be cast. As a trade-off, a longer vector can represent more complicated keywords but brings more overhead. On the other hand, a too short vector would make information loss normal, which brings severe hurt to representation rationality.
 
 After encoding, we have vectors of uniform length representing keywords. The position of each letter in the original keyword is encoded in these vectors, making a critical difference with traditional methods.

 \subsubsection{Infect}

 A simple insight to realize fuzzy search is to raise the tolerance for letter dislocation when transforming words to a standardized representation (such as the uni-gram vector adopted in our scheme). To the tolerance of the produced uni-gram vector, we propose an 'Infect' mechanism, which makes the most difference between our proposed method and previous methods. Each nonzero bit of the vector after ``Encode'' would share its weights with neighboring bits and the relation is determined by an \emph{Infection Function}. After \emph{Infection}, bits of original representation vector may be transformed from binary to float. A typical kind of \emph{Infection Functions} can be formulated as:

 \begin{equation}
	\delta w(d) = 
	\begin{cases}
		\frac{1}{s^{d/26}} & d \leq 26u , d\%26 = 0\\
		0 & otherwise\\
	\end{cases}
	\label{infection}
\end{equation}
where $d$ is the bit distance between two bits, $s$ is a factor to adjust the infection strength. Note $u$ is another to adjust how far the infection can spread and the farthermost distance is $d_{max}$ = $26u$. Because same letters on the neighboring position of the original keyword are mapped into two 26-bit-far-away bits, only when $d\%26=0$, infection happens. Such a mechanism would weaken the negative effect brought by letter dislocation.

To explain that in detail, we study it with an example. A keyword ``add'' is already encoded into a OPU vector after ``Decompose'' and ``Encode''. If we set $s=2$ and $u=2$, \emph{Infection} happens: $V_{KW}[1], V_{KW}[30]$ and $V_{KW}[56]$ are 1, others being 0. In the first wave of infection, value of $V_{KW}[27], V_{KW}[4], V_{KW}[56], V_{KW}[30]$ and $V_{KW}[82]$ increases by $\delta w = \frac{1}{2^1} = 0.5$. Then the second wave of infection follows: $V_{KW}[53], V_{KW}[82], V_{KW}[4]$ and $V_{KW}[108]$ are further increased by $\delta w = \frac{1}{2^2} = 0.25$. So far, infection finishes because $s=2$. Finally the OPU vector generalized from the keyword ``add'' becomes $V^{\prime}_{KW}$ with $u=2$:

\begin{equation}
	V^{\prime}_{KW}[i] = 
	\begin{cases}
		1.5 & i = 30, 56 \\
		1 & i = 1 \\
		0.75 & i = 4, 82 \\
		0.5 & i = 27 \\
		0.25 & i = 53, 108 \\
		0 & otherwise
	\end{cases}
	\label{unigramfinal}
\end{equation}

\subsubsection{Analysis of Order-preserved Unigram (OPU)}
We analyze the improvement of the proposed OPU over previous 'bi-gram'\cite{firstfuzzy} and 'uni-gram'\cite{target} design in this part. Comparison is performed through examples with different requirements for fuzzy search taken into consideration. The similarity of two keywords is simply quantified in Euclidean distance as:

\begin{equation}
	Dist[V_1][V_2] = \sqrt{\sum_{i=1}^{N}{(V_1[i] - V_2[i])^2}},
\end{equation}

\noindent where $V_1$ and $V_2$ are respectively the representation vector of two keywords. $N$ is the length of vectors. In fuzzy search, we want simple typo brings no severe decrease to similarity score. In other words, if $V_1$ is the correct form of a keyword and $V_2$ is a corresponding fuzzy form, their distance should be as short as possible so that they are considered to be largely similar during a search. On the other hand, if $V_1$ and $V_2$ are different wanted keywords and they are similar under some metric, a robust representation design can still distinguish them after vector representation. For instance, ``add'' and ``dad'' are not expected to be thought ``similar'' or even ``the same'' under a good scheme, which may bring huge bias into search results.

Considering various requirements for fuzzy search, three types of misspellings cases should be taken into consideration:

\paragraph{Letter misspelling}
Letter misspelling indicates when letters in a word are replaced by some incorrect ones. For example, ``beer'' can be misspelled as ``berr''.

\paragraph{Wrong letter order}
Wrong order of letters indicates when words are consist of a uniform set of letters, but some letters in them are arranged with the wrong order. For example, the word ``bere'' may be typed as ``beer'' for wrong letter order.

\paragraph{Insertion/Absence of letter}
Insertion or absence of letter(s) in a word occurs frequently as well, causing typos in the text. For example, the word ``pen'' may be misspelled to be ``pean'' or ``pn'' in this case. 

An ideal keyword decomposing approach should recognize the high similarity between a word and its typos. On the other hand, as an obvious trade-off, when a typo suffers from the severe difference from the correct form, the similarity should no longer be high, or the approach would become invalid to distinguish some different but similar words. For the three listed fuzzy cases, we calculate the relevance scores under different word decomposing approaches. The results are shown with examples in TABLE \ref{casecompare}. 

\begin{table}
\renewcommand\arraystretch{1.5}
	\centering
	\begin{tabular}{c|c|c|p{0.65cm}|p{0.65cm}|c}
		\toprule
		Fuzzy type & Correct & Wrong & Bi-gram & Uni-gram & OPU \\
		\hline
		\multirow{2}*{Misspelling} & add & aad & \centering2 & \centering2 & $\sqrt{3.125}$ \\
		 ~ & bear & beer & \centering2 & \centering2 & $\sqrt{3.25}$ \\
		 \hline
		 \multirow{2}*{Wrong Order} & used & uesd & \centering$\sqrt{6}$ & \centering0 & $\sqrt{1.75}$ \\
		 ~ & pear & paer & \centering4 & \centering0 & 1 \\
		 \hline
		 \multirow{2}*{Insertion/Absence} & pen & pn & \centering$\sqrt{3}$ & \centering1 & $\sqrt{3.25}$ \\
		 ~ & pen & pean & \centering$\sqrt{3}$ & \centering1 & $ \sqrt{2.75}$ \\
		\bottomrule
	\end{tabular}
	\caption{The comparison of similarity scores under different schemes. The similarity is represented by Euclidean distance and parameters for OPU are set as: $u=2$ and $s=2$.}
	\label{casecompare}
\end{table}

Except for the three basic types of misspellings, OPU handles many other cases better than 'uni-gram' and 'bi-gram'. The most obvious shortcoming of traditional 'uni-gram' is that only the composition of letters in a word is recorded after decomposing, the information of their position is lost totally. Such a shortcoming makes it unable to distinguish different words in many cases. For example, for the traditional 'uni-gram' mechanism, anagrams like ``listen'' and ``silent'' produce the same keyword vector after transformation, which is incapable of satisfying users' demand. For example, it is apparently unacceptable to regard ``Dad is silent'' and ``Dad is listen'' as the same queries. While in our proposed OPU, the position information is also encoded into the final word representation, thus anagrams could no longer compromise our approach.

In summary, based on 'uni-gram', our proposed OPU not only inherits all its advantage but also encodes the position of letters in the original word into final representation, which enhances our scheme in many cases. We qualitatively compare the ability of three mentioned mechanisms to show the difference between different words in different cases and the result is shown in TABLE \ref{schemecompare}. Note that ``Ex-1'' concludes the case with non-alphabet involved in a string.

\begin{table}[!htb] 
\renewcommand\arraystretch{1.5}
 \centering
 \begin{tabular}{c|c|c|c|c}
 \toprule[1pt]
    Keyword \#1 & Keyword \#2 & bi-gram & uni-gram & OPU\\
    \hline
    listen & silent & Yes & No & Yes\\
    \hline
    2case & case2 & No & No & Yes\\
    \hline
    exams & exam & Yes & Yes & Yes\\
    \hline
    running & run & No & Yes & Yes\\
    \hline
    useful & use & No & Yes & Yes\\
    \hline
    Ex-1 & Ex-1 & No & Yes & Yes\\
  \bottomrule[1pt] 
 \end{tabular}
 \caption{Comparison among three schemes}
 \label{schemecompare}
\end{table}

\begin{figure}
	\centering
	\includegraphics[width=\linewidth]{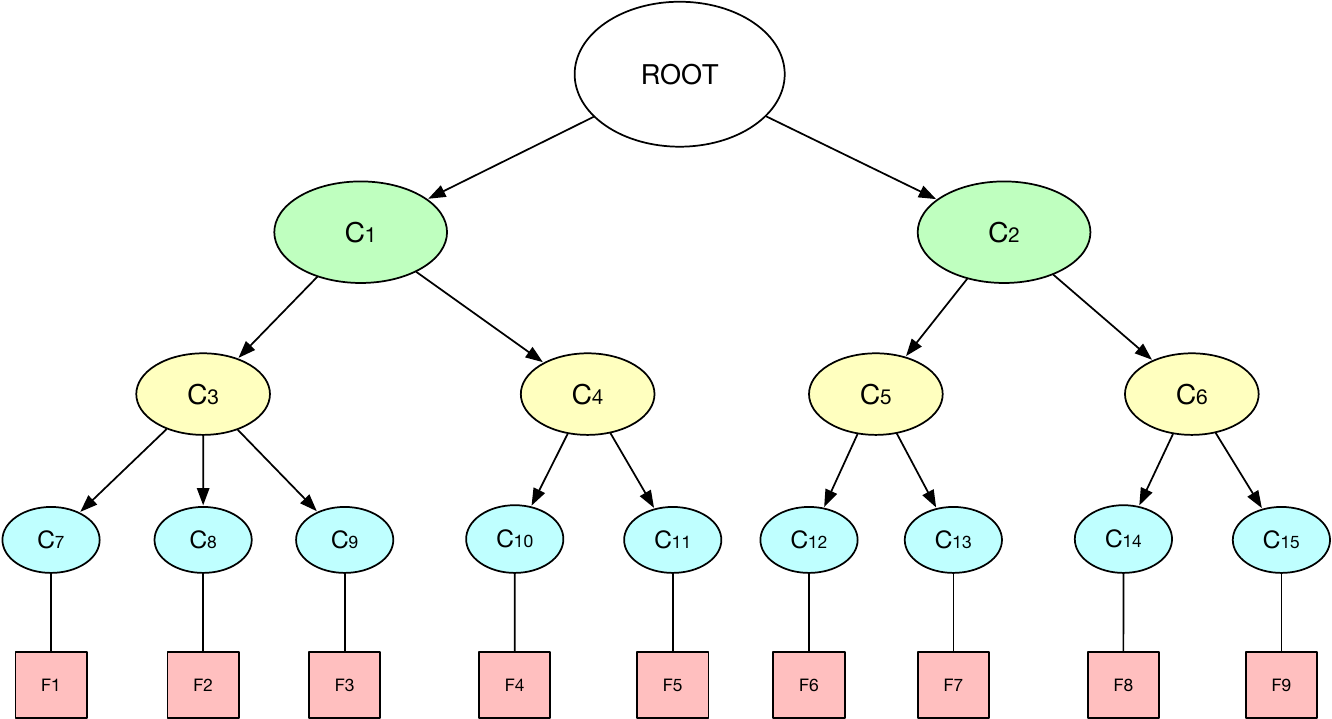}
	\caption{Overview of Hierarchical Index Tree}
	\label{tree}
\end{figure}

\begin{figure*}  
\centering
\includegraphics[width=\linewidth]{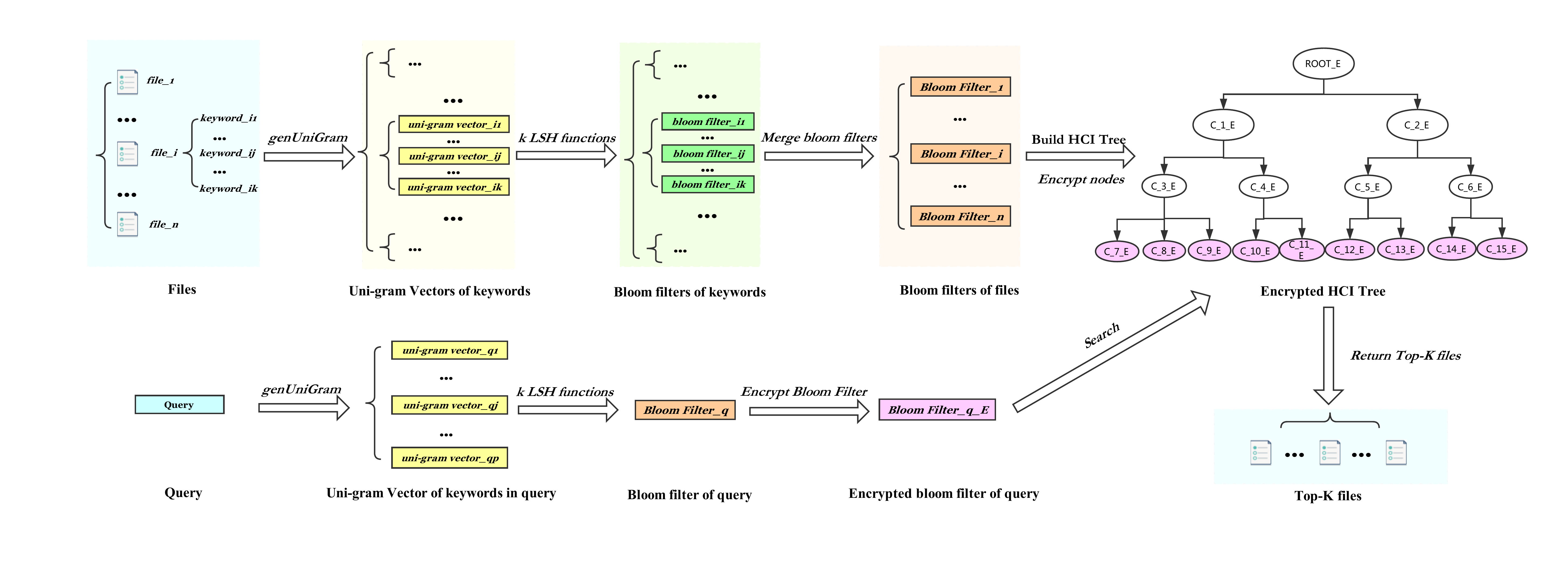}  
\caption{Overview of the complete procedure of proposed scheme}  
\label{general process}  
\end{figure*} 

\subsection{``Hierarchical Index Tree'' (HIT)}
As the other main contribution, we exploit an efficient data structure to organize the file indices. Instead of the most naive linear organization, we design a tree-based index organization. To be precise, we build a hierarchical index tree to organize the file indices for faster file search. To relieve computation overhead, we do not index a file straightforward through its full keyword vector, which usually is of high dimension. Instead, we map original word vectors into some intermediate representation of lower dimension and then perform indexing on it. As mentioned before, we choose ``bloom filter'' as the intermediate representation.

\subsubsection{Construction of HIT}
Given the file representation vectors, each of which encodes the keyword information of a file, we propose HIT to organize them. To build HIT, we need to divide bloom filters into clusters. Some nearby clusters may form a larger cluster standing on a higher layer in HIT. We propose an improved dynamic K-means Algorithm to plan these clusters of different levels, through linking which we could build the final HIT. We explain the algorithm in Algorithm \ref{cluster}. Instead of giving a fixed tightness factor to determine the final cluster number, we compare the average point distance and minimum point distance in a cluster to determine whether this cluster should be further subdivided.

To generalize clusters of higher level, the clusters of lower-level are represented by their center coordinates and thus regarded as ``points''. Algorithm \ref{cluster} is therefore capable of clustering some tiny clusters into a larger one. Such operation is kept performing until a final cluster containing all points is generalized, and the construction of HIT is done.

\begin{algorithm}
\caption{HIT Construction}
\label{cluster}
{\bf Input:}\\
1. $\mathcal{P} = \{P_1, P_2,..,P_m\}$: the set of points to be clustered\\
2. $e$: a factor to adjust the desired tightness of clusters\\
3. $D$: function to calculate the distance of two points in $\mathcal{P}$\\
{\bf Output:}\\
1. $\mathcal{C} = \{C_1, C_2,...,C_n\}$: coordinates of built cluster centers\\
2. $\mathcal{L} = \{L_1, L_2,...,L_m\}$: cluster index of each point
\begin{algorithmic}[1]
\State $n \leftarrow 1$
\State $Stable \leftarrow False$
\While{not Stable}
	\State Divide $\mathcal{P}$ into $n$ clusters through K-means method
	\State Update $\mathcal{C}$ and $\mathcal{L}$
	\State $Stable \leftarrow True$
	\For{$C_i$ in $\mathcal{C}$}
		\State Calculate the distance matrix: $\mathcal{M}^{i}_{uv} = D(P_u, P_v)$, where $P_u, P_v \in C_i$
    	\State $R_{avg} \leftarrow$ the average distance in $M^{i}$
    	\State $R_{min} \leftarrow$ the minimum distance in $M^{i}$
    	\If{$R_{min} < eR_{avg}$}
    		\State $n \leftarrow n + 1$
    		\State $Stable \leftarrow False$
    	\EndIf
	\EndFor
\EndWhile
\Return $\mathcal{C}$ and $\mathcal{L}$
\end{algorithmic}
\end{algorithm}

\subsubsection{Search in HIT}
To search files in HIT, we need to calculate the relevance score between the representation vectors in HIT nodes and a given query vector, which usually is generalized from a query string. Only the leaf nodes in HIT represent real files and search process is expected to return $k$ most similar stored files. To increase time efficiency, an intelligent search algorithm should not calculate the relevance score of every leaf between the target vector. Therefore, how to find reliable top-$k$ most similar files without too much computation is the core problem. We thus design a search algorithm adapted to the HIT structure as explained in Algorithm \ref{search_algorithm}, which could be notated as $Search(R, T, k, Rs)$.

Only to traverse all files can ensure that the literally ``top-k most relevant files'' are always found. Linear traversing requires a time complexity of $\mathcal{O}(n)$, where $n$ is the scale of stored files, which is unacceptable in most cases. Through the proposed search algorithm, we try to find a proper trade-off between search accuracy and time efficiency. In other words, we look up required $k$ files with $\mathcal{O}(log_n)$ time complexity and returned $k$ files can be expected to be included in the global ``top-k most relevant files''. The designed experiment proves that HIT improves time efficiency to a high extent without bringing much damage on search accuracy.

\begin{algorithm}
\caption{Search in HIT}
\label{search_algorithm}
{\bf Input:}\\
1. $R$: root node of current HIT\\
2. $T$: the representation vector of input query\\
3. $k$: the number of files to be returned\\
4. $Rs(A, B)$: function of relevance score between $A$ and $B$\\
{\bf Output:}\\
$\mathcal{F} = \{F_1,...,F_k\}$: found top-k similar files
\begin{algorithmic}[1]
\While{children of $R$ are not leaves}
    \State children nodes of $R$ are $C(R) = \{C_1,...,C_p\}$
    \For{$i \leftarrow 1$ to $p$}
        \State $S_i \leftarrow Rs(C_i, T)$
    \EndFor
    \State find $S_k \leftarrow max\{S_i | 1 \le i \le p\}$
    \State $R \leftarrow C_k$
\EndWhile
\State children of $R$ are $C(R) = \{C_1,...,C_q\}$
\State $\mathcal{S}=\{S_1,...,S_p\}$ where $S_i = Rs(C_i, T)$
\State $R^\prime \leftarrow R$
\If{$k \le q$}
    \State $\mathcal{C}=\{C^\prime_1,...,C^\prime_k\} \leftarrow$ points ranking top-k in $\mathcal{S}$
    \State $\mathcal{F}=\{F_1,...,F_k\}$: files of $C^\prime_i(1 \le i \le k)$
\Else
    \While{$|C(R^\prime)|=1$ or $| L(R^\prime) | < k$}
        \State \% {$L(R^\prime)$ returns number of leaves under $R^\prime$}
        \State $R^\prime \leftarrow$ father of $R^\prime$
    \EndWhile
    \State $\mathcal{F}_1 \leftarrow$ files corresponding to points in $C_R$
    \State $\mathcal{F}_2 \leftarrow Search(R^\prime, T, k-p, Rs)$
    \State $\mathcal{F} = \mathcal{F}_1 \cup \mathcal{F}_2$
\EndIf
\Return $\mathcal{F}$
\end{algorithmic}
\end{algorithm}

\begin{figure}
	\centering
	\includegraphics[width=3.0in]{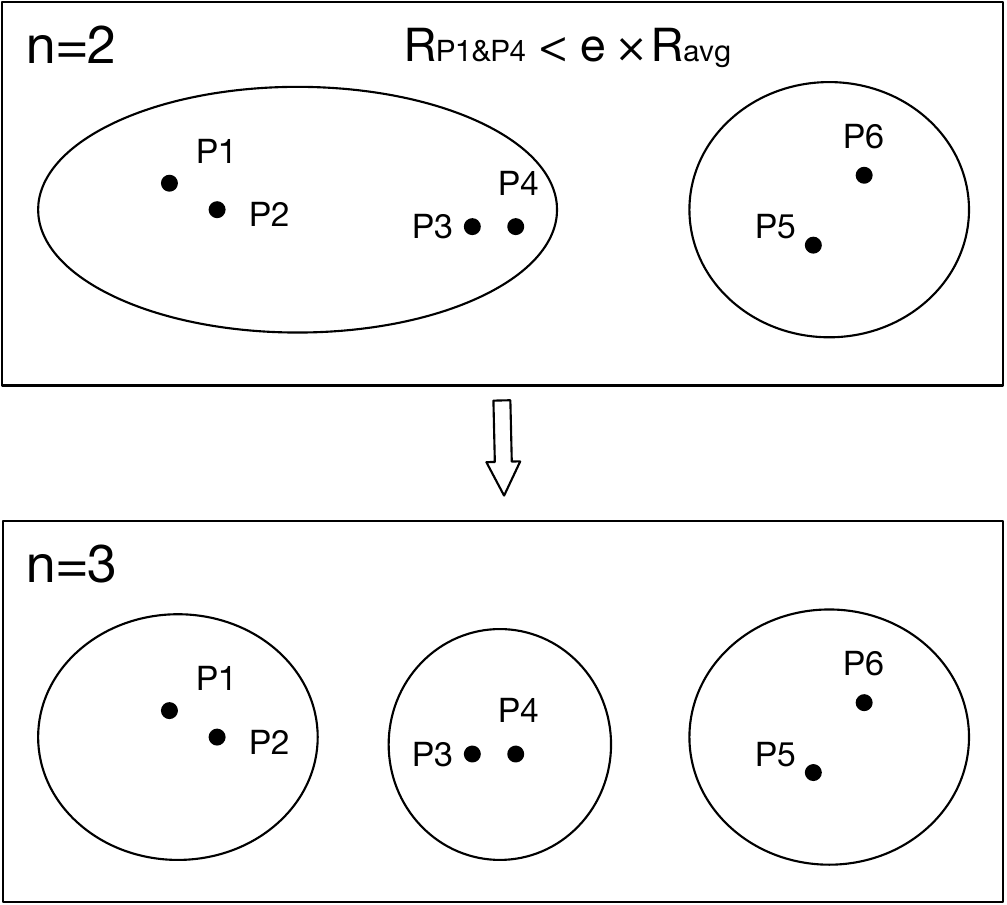}
	\caption{Adjust cluster number in Algorithm \ref{cluster}}
	\label{adjustn}
\end{figure}

\section{Architecture Construction}
In this section, we explain the complete pipelines of our proposed architecture in detail. Note that the scheme comprehensively leverages multiple algorithms and data structures in different steps dependent on each other. However the general architecture is still flexible with sub-modules replacement. For example, replacing bloom filter with other data structure or adopting a different encryption approach would not disable the architecture but only change fine-grained operations. The overview of the proposed architecture is visualized in Figure \ref{general process} in detail.

\subsection{Keyword Extraction and Preprocessing}
Given the set of files to be outsourced as $\mathcal{F}=\{F_1,F_2,...,F_n\}$, the first step is to extract the keywords from them. Preposition, pronoun, auxiliary words and other keywords without concrete semantic meaning are filtered out. Remaining keywords are noted as $\mathcal{K}=\{\mathcal{K}_1,...,\mathcal{K}_n\}$, where $\mathcal{K}_i$ is the set of keywords extracted from $F_i$. We have $\cup^{n}_{i=1}\mathcal{K}_i = \{KW_1,...,KW_N\}$. Then, we apply Stemming Algorithm\cite{porterstemmer} on these keywords to eliminate grammatical diversity. For example, ``listening‘’, ``listen‘’ and ``listened‘’ would be all stemmed into a uniform keyword 'listen'. After stemming, we then calculate the ``term frequency'' and ``inverse document frequency''(TF-IDF)\cite{TFIDF} value of each keyword for further use. 

Considering there are in total $N$ legal keywords, the TF value of the $i^{th}$ keyword $KW_i$, in the $j^{th}$ file $F_j$, is formulated as:
\begin{equation}
    TF_{i,j} = \frac{n_{i,j}}{\sum^{N}_{k=1}n_{k,j}},
\end{equation}
where $n_{i,j}$ is the frequency of $KW_i$ in $F_j$. The IDF value is on the other calculated as:

\begin{equation}
    IDF_{i} = \frac{|\mathcal{F}|}{|\{F_k | KW_i \in F_k\}|}
\end{equation}
The TF-IDF value of $KW_i$ in the file $F_i$ is thus simply calculated by:

\begin{equation}
    TFIDF_{i,j} = TF_{i,j} \times IDF_{i}
\end{equation}

\subsection{Generalization of representation vectors of files}
To relieve computation overheads, we would transform the original keywords in natural language form into some numerical representation. For computation and storage convenience, they are in general transformed into vectors of uniform length. Then we build a structured representation of each file based on the representation vectors of all contained keywords in it. In our proposed scheme, we follow two steps to reach the goal.
\subsubsection{Keyword representation}
To represent a keyword in a uniform numerical form, we transform each keyword into a corresponding OPU vector as introduced in the aforementioned sections. Then, with $l$ LSH functions, $\mathcal{H}=\{H_1,H_2,...,H_l\}$, a keyword would be encoded into at most $l$ bits in a bloom filter with else bits all zero.

\subsubsection{File representation}
All keywords in a file haven been represented already with bloom filter and the file can thus be represented a set of keyword bloom filters. For example, for a file $F_i$, we have $BF_{F_j} = \{bf_{1,j},...,bf_{u,j}\}$ where $bf_{k,j}$ is the bloom filter of the $k^{th}$ keyword in $F_j$. Furthermore, we can represent the file in a bloom filter as well. The bloom filter of a file is of the same length with that of keywords. Through bit-wise addition and pre-calculated TF-IDF value of each keyword, the final representation of $F_j$ is formulated as:

\begin{equation}
    bf_{F_j} = \sum^{u}_{k=1} TFIDF_{k,j} \times bf_{k,j}
\end{equation}

Till now, we've got the final representation of files to be outsourced. For queries, which would be involved in the architecture in the search stage, we transform them into a numerical representation as same as the process for files.

\subsection{Construction of encrypted HIT}
For search efficiency, we would outsource an aforementioned HIT for file indexing to the remote servers. Besides, for privacy protection, we have to encrypt built HIT before the outsourcing. In this section, we introduce the process to build and encrypt the HIT.

\subsubsection{Building HIT}
Before this step, each file has been represented in a bloom filter, which can also be regarded as a point in a $l_b-dimensional$ space, where $l_b$ is the length of adopted bloom filters. As introduced in the aforementioned section and illustrated in Algorithm \ref{cluster}, we can build an HIT to organize the indices of files now.

\subsubsection{Encryption of HIT and query}
To ensure security and privacy, simply to transform data into bloom filters may be inadequate because such a transformation is determinative. Hence, we adopt secure kNN\cite{song2000} algorithm to encrypt all nodes in the HIT as well. On the other hand, a query will also be encrypted in a corresponding manner to enable the calculation of relevance between files and query. This process is decomposed into steps as follows:

\paragraph{KeyGen(m)}
Given a security parameter $m$, this method generates a security key $SK$, which is a tuple $(M_1, M_2, S)$. $M_1$ and $M_2$ are both $m$-dimensional invertible matrices and $S \in {(0,1)^m}$ is an $m$-dimensional vector.

\paragraph{EncIndex($I, SK$)}
With secure kNN to encrypt an index vector $I$, which should be a bloom filter in our scheme and of length $m$, it should be first split into two vectors $(I^\prime, I^{\prime \prime})$  as follows:

\begin{equation}
	I^\prime[j] = 
	\begin{cases}
		I[j], S[j] = 1 \\
		\frac{1}{2} \times I[j] + r, S[j] = 0
	\end{cases} , 1 \leq j \leq m
\end{equation}

\begin{equation}
	I^{\prime \prime}[j] = 
	\begin{cases}
		I[j], S[j] = 1 \\
		\frac{1}{2} \times I[j] - r, S[j] = 0
	\end{cases} , 1 \leq j \leq m
\end{equation}
where $r$ is a small randomness introduced for security consideration. Finally, the encrypted expression of $I$ is $T_F=\{M^{T}_1 \cdot I', M^{T}_2 \cdot I''\}$.

\paragraph{EncQuery($Q, SK$)}
For queries in the search stage, it should also be encrypted to prevent information leak. Following the previous steps, a query has also been represented in an $m$-dimensional bloom filter, which is notated as $Q$. To encrypt it into a trapdoor, symmetric operations are operated to split it into two vectors $(Q^\prime, Q^{\prime \prime})$ as well:

\begin{equation}
	Q^\prime[j] = 
	\begin{cases}
		Q[j], S[j] = 0\\
		\frac{1}{2} \times Q[j] + r, S[j] = 1
	\end{cases} , 1 \leq j \leq m
\end{equation}

\begin{equation}
	Q^{\prime \prime} =
	\begin{cases}
		Q[j], S[j] = 0 \\
		\frac{1}{2} \times Q[j] - r, S[j] = 1
	\end{cases} , 1 \leq j \leq m
\end{equation}

Finally the trapdoor of the query vector is expressed as: $T_Q=\{M^{-1}_1 \cdot Q', M^{-1}_2 \cdot Q'' \}$

\subsection{Search}
While once all data encrypted, we need to re-introduce the calculation of relevance score. Fortunately, secure kNN has a great feature that it allows invariant relevance score through naive inner production. Given an encrypted file index $T_F$ and an encrypted query $T_Q$, we can obtain their relevance score as follows:

\begin{equation}
\label{knncal}
\begin{aligned}
	&Rs(T_F, T_Q) \\
	&= T_F \cdot T_Q\\
	&= \{M^{T}_1 \cdot I', M^{T}_2 \cdot I''\} \cdot \{M^{-1}_1 \cdot Q', M^{-1}_2 \cdot Q'' \}\\
	&= I' \cdot Q' + I'' \cdot Q''\\
	&= \sum_{j=1}^{m}(I'[j] \cdot Q'[j] + I''[j] \cdot Q''[j])\\
	&= \sum_{j=1, S[j]=0}^{m} (\frac{1}{2}I[j] + r) \cdot Q[j] + (\frac{1}{2}I[j] - r) \cdot Q[j]\\
	&+ \sum_{j=1, S[j]=1}^{m} (\frac{1}{2}Q[j] + r) \cdot I[j] + (\frac{1}{2}Q[j] - r) \cdot I[j]\\
	&= I \cdot Q
\end{aligned}
\end{equation}

Eq \ref{knncal} proves that inner production on two vectors can produce the same result before and after the encryption by secure kNN. So far, we can search for the top-$k$ most relevant files through HIT referring to a given query. The required algorithm has been introduced in previous sections and concluded in Algorithm \ref{search_algorithm}.

\subsection{Verification}
Similar as explained in \cite{HCandVeri}, thanks to the construction of HIT, returned files can be further verified to confirm the correctness, completeness, and freshness. To support this mechanism, the data owner needs to build a signed tree in advance and outsource it to the cloud server together with the index tree. Once the search finishes, the cloud server returns top-$k$ files together with the \emph{signed sub-tree} along the search path to the data user. For example, referring to Fig \ref{tree}, if files in $C_7, C_8, C_9, C_{10}$ are returned as search result, the representation vectors stored in $C_1, C_3, C_4, C_7, C_8, C_9, C_{10}$ would be all returned as well because these nodes are gone through in search path. Here, we could adopt a signature algorithm such as $RSA$\cite{rsa} to generate this signed hash tree. Then, with the received \emph{signed sub-tree}, the data user calculates the signature of each node with its corresponding representation vector and compares the calculated signature with the returned signature. The returned files pass the test, and as claimed in \cite{HCandVeri}, their correctness, completeness, and freshness could be verified without considering the real-time data update on remote servers.

\section{Privacy Analysis}
In this section, we state the background settings and analysis of the security promise of our proposed architecture theoretically. The analysis is performed under two threat models, where privacy leakage resulting from untrusted data use is out of the discussion. The authorization of data users and remote servers is also not in the scope of this section. We assume all data users taken into consideration are already certified by the data owner.

\subsection{Threat Model}
The cloud server is considered as honest-but-curious\cite{honestbutcurious}, which means that it receives queries and executes the search as commanded, but it tries to derive sensitive information from queries and stored encrypted data at the same time. The requirement of privacy in our scheme is as same as that defined in\cite{firstfuzzy,target,securitydefine}. Thus, there are two threat models asking for different levels of privacy protection:
\paragraph{Known Ciphertext Model}
In this threat model, the cloud server is expected to gain only the content of encrypted files, encrypted file index and encrypted trapdoor of received queries.
\paragraph{Known Background Model}
In this threat model, the cloud server is also interested in the statistical background information of encrypted data. It could perform a statistical attack to obtain more information about keywords\cite{cao2014,statattack} involved in the search. Attach under this threat model also reveals unencrypted information through collecting and inference.

\subsection{Security Objectives}
Because we adopt the encryption-before-outsourcing schemes\cite{encryptbeforeoutsource} in our proposed architecture, the privacy of file content is already guaranteed unless adopted encryption is broken. Then, there are other security objects that we need to take into consideration:

\paragraph{Keyword Unvisiability}
File index and trapdoor of queries are both transformed from extracted keywords. The real content keywords should also be protected from being known by remote servers.
\paragraph{Trapdoor unlinkability}
Remote servers should be prevented from learning the content of query/files from repeated search tasks. Thus, trapdoor generation from query/files should be fully deterministic. Thus, the same query string or file should not be always transformed into the same trapdoor.

We adopt bloom filter structure and secure kNN to ensure security and privacy. Because the generalization of bloom filters is deterministic with a certain family of LSH functions, it cannot reach the requirement of 'Trapdoor unlinkability'. However, during the encryption with secure kNN, randomness is introduced into all trapdoors, which helps our scheme to be capable of these security and privacy criteria. Detailed proof can be simply borrowed from that provided in\cite{song2000, sun2013privacy}.

\section{Experiments and Evaluation}
In this section, we design groups of experiments to estimate the performance of our scheme. We use the real-world natural language dataset '20NewsGroups'\cite{20newsgroup} as the raw plaintext file source. Main programs are written in Python2.7. All experiments were performed on a server having an 'Intel(R) Xeon(R) CPU E5-2680 v4 @ 2.40GHz' and 16GB available DDR4 memory space. All experiments are run on Ubuntu 16.04 LTS operation system.

\subsection{Evaluation Metrics}

Since parameters setting matters a lot in experiments, we specify it at first. Unless declared otherwise, involved paramerts are set by default as:

\begin{itemize}
    \item $s=2, u=2$ for ``Infection'' stage in generalizing OPU
    \item each query contains 5 keywords
    \item bloom filters have in total 8000 bits
    \item in total $l=20$ LSH functions are used to build bloom filters
    \item each query asks for 20 returned files
    \item $e$ in Algorithm \ref{cluster} is set to be 0.4
\end{itemize}

Each text file in the dataset '20newsgroup' has a set of keywords whose scale varies a lot. To relieve the negative effect brought by this bias, we first filter the dataset by referring to how many keywords a file contains. We select 3583 files from '20newsgroup', each of which has at least 200 keywords and at most 400 keywords. In total 41558 raw keywords are extracted from these selected raw files. After stemming, keywords still sum to more than 3000. Instead of simply setting a threshold of relevance score to calculate search accuracy, we introduce a more practical accuracy metric for evaluation, which named \emph{overlap rate of top-k files}. Because output return top-k files affect what the data user gets from his query, this metric is expected to be better estimated the search accuracy in data users' real experience. The accuracy under such metric is formulated as: 

\begin{equation}
	Accuracy = \frac{| \{ f | f \in FP_k, f \in FE_k \} |}{k},
\end{equation}
where $FP_k$ is the set of top-k files returned through traversing search on plaintext files, and $FE_k$ is the set of top-k files returned through encrypted search. For instance, given the same query, the search performed on encrypted data and plaintext search return the same set of top-k files. Then $Accuracy$ reaches its upper limit, i.e., $Accuracy = 1$. When $k=3$, if the plaintext search returns a collection of files $FP_k=\{F_1, F_3, F_5\}$ and the search on encrypted data returns $FE_k=\{F_1, F_3, F_6\}$, the accuracy should be $\frac{2}{3}$. We compare our proposed scheme with the most popular scheme based on ``uni-gram'' keyword decomposing\cite{target}.

\subsection{Non-fuzzy Search}
\begin{figure}[htb]
	\centering
	\includegraphics[width=\linewidth]{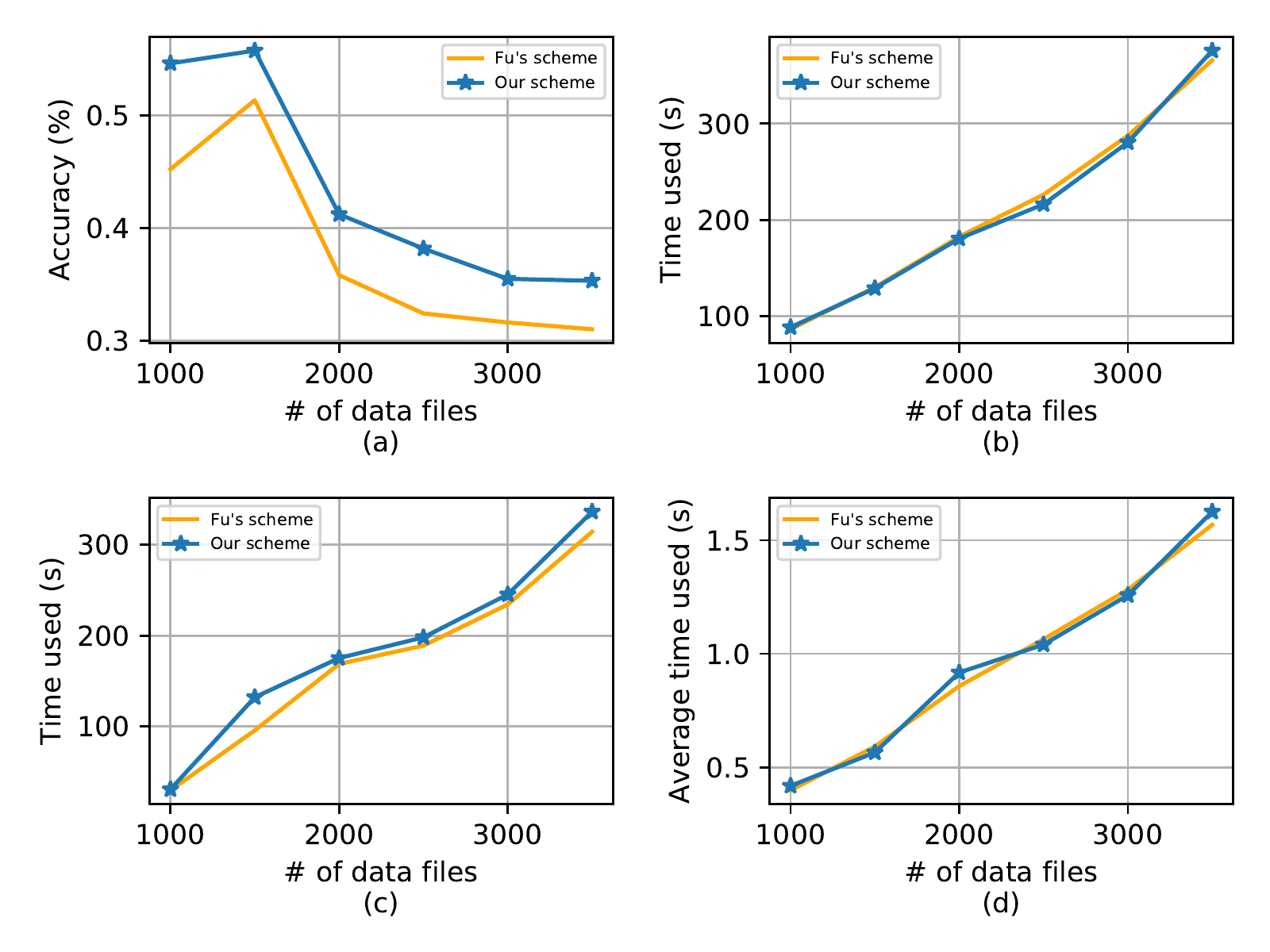}
	\captionof{figure}{Metrics under different volumes of dataset. a): accuracy of returned result; b): time used to generalize dictionary from raw dataset; c): time used to generalize bloomfilters of files; d): average time consumed to perform a query}
	\label{filenum}
\end{figure}

As a basic requirement, fuzzy search approaches should definitely show convincing performance when no typo or other misleading information is involved. We test the search accuracy and time efficiency on text file datasets of different volume. The experiment result is shown in Figure \ref{filenum}. The search time for both of the schemes are tested under the naive linear search for variable control.

For the different volume of the dataset, our proposed scheme achieves higher accuracy compared with Fu's scheme\cite{target} with almost the same time performance. The main difference between these two schemes is how to represent a keyword, based on ``uni-gram'' of ``order preserved uni-gram''. Hence, the experiment proves that the introduction of OPU in our proposed scheme does not bring obvious extra computation overheads but it contributes to search accuracy to an evident extent.

\begin{figure}[!htb]
	\centering
	\includegraphics[width=\linewidth]{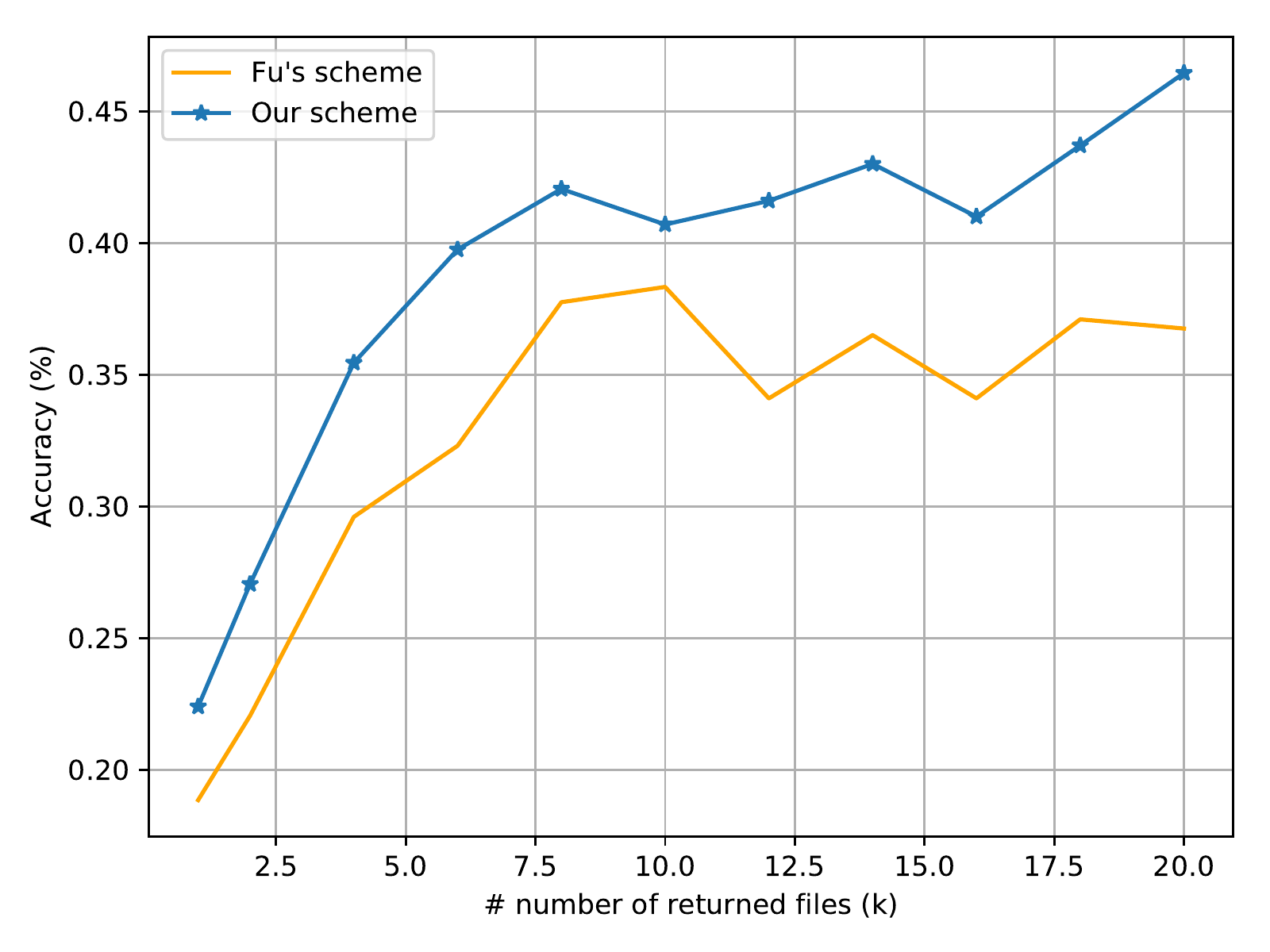}
	\captionof{figure}{Accuracy with different numbers of returned files (different values of $k$)}
	\label{valuek}
\end{figure}

In practice, the number of returned files, $k$, ought to be allowed fluctuant and adjusted by the data user. Therefore, a robust search scheme should achieve steady search accuracy when $k$ changes. We control the value of $k$ and maintain other variables unchanged in the experiment for comparison. The result is shown in Figure \ref{valuek}. Generally speaking, along with the increase of $k$ (return files for a query), search accuracy increases and our scheme always achieves higher accuracy. 

\begin{figure}[!htb]
	\centering
	\includegraphics[width=\linewidth]{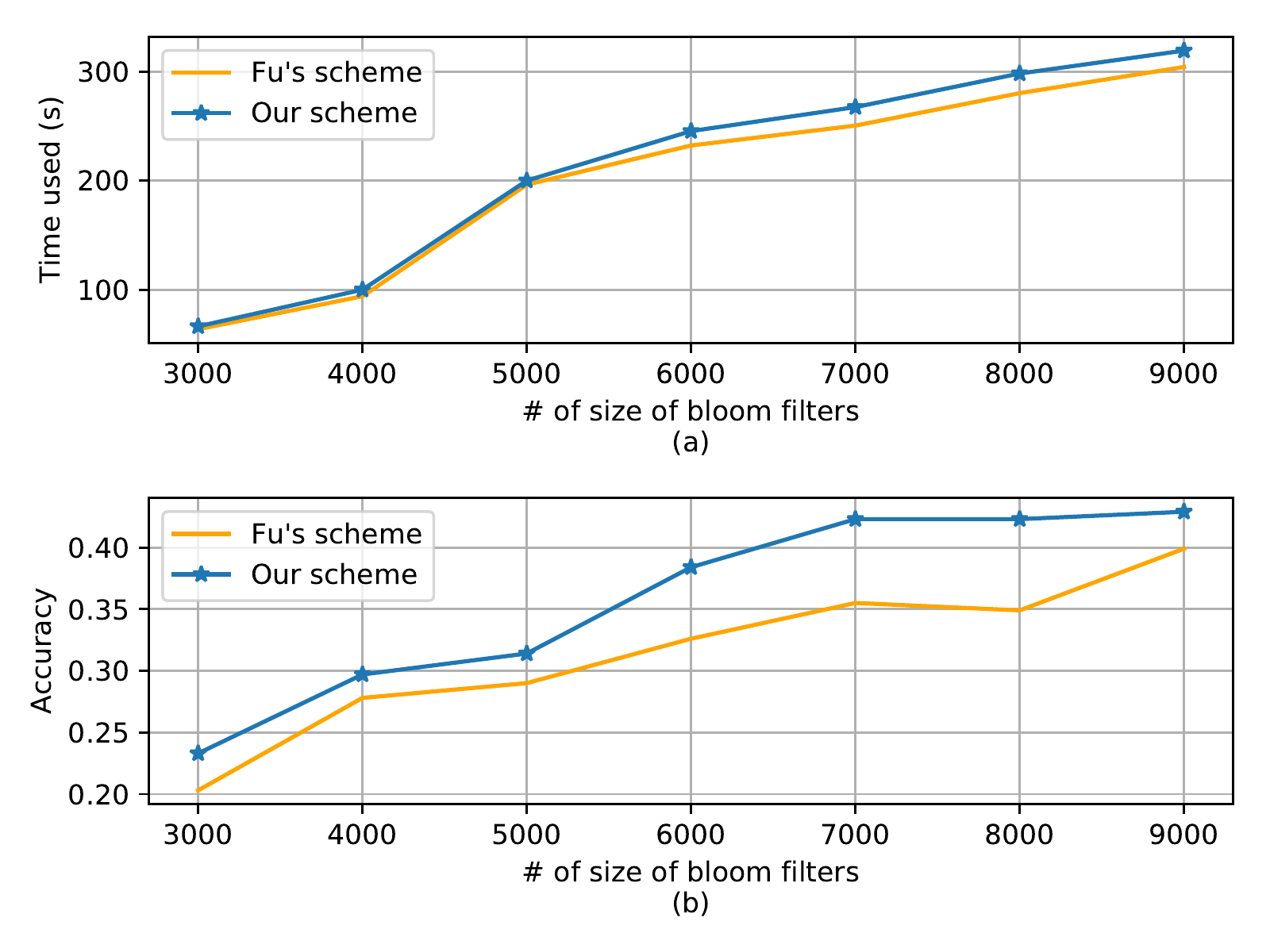}
	\captionof{figure}{Influence of the size of bloom filter ($Len_{bf}$) on the accuracy and time usage}
	\label{bfsize}
\end{figure}

Similar to other searchable encryption mechanisms\cite{target,bellovin2004privacy,chang2005privacy} based on Bloom Filter\cite{bloomfilter}, another critical trade-off between time efficiency and accuracy is the size of bloom filter. A bigger bloom filter decreases the chance of hashing collision but it obviously increases the computation and memory storage overhead. To evaluate the influence of bloom filter size on our scheme, we set another group of experiments and the result is shown in Fig \ref{bfsize}. Referring to the experiment, we find that our proposed scheme performs better in the set range of bloom filter size. The time efficiency is nearly the same as in that in \cite{target}. On the other hand, the increased length of bloom filter brings no extra time consumption compared. It is because that OPU introduces no extra computation overheads into the bloom filter generalization stage.

\begin{figure}[!htb]
	\centering
	\includegraphics[width=\linewidth]{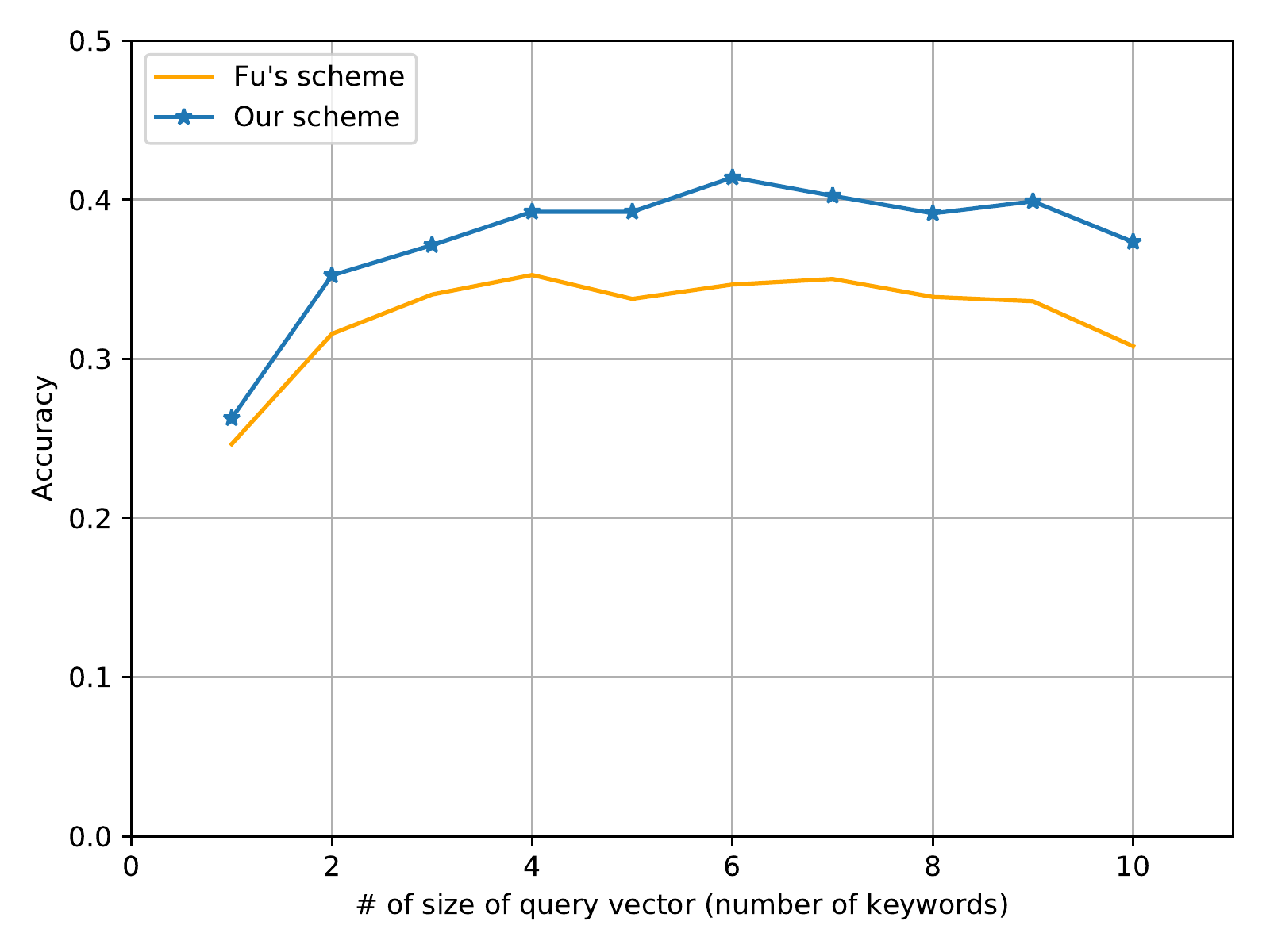}
	\captionof{figure}{Search accuracy with different size of query string ($Len_{query}$)}
	\label{querysize}
\end{figure}

At last, we also evaluate performance with different lengths of queries. The result is reported in Fig \ref{querysize}. Our proposed scheme performs steadily with different queries consisting of 1-10 keyword(s).

\subsection{Fuzzy Search}

\begin{figure}[!htb]
	\centering
	\includegraphics[width=\linewidth]{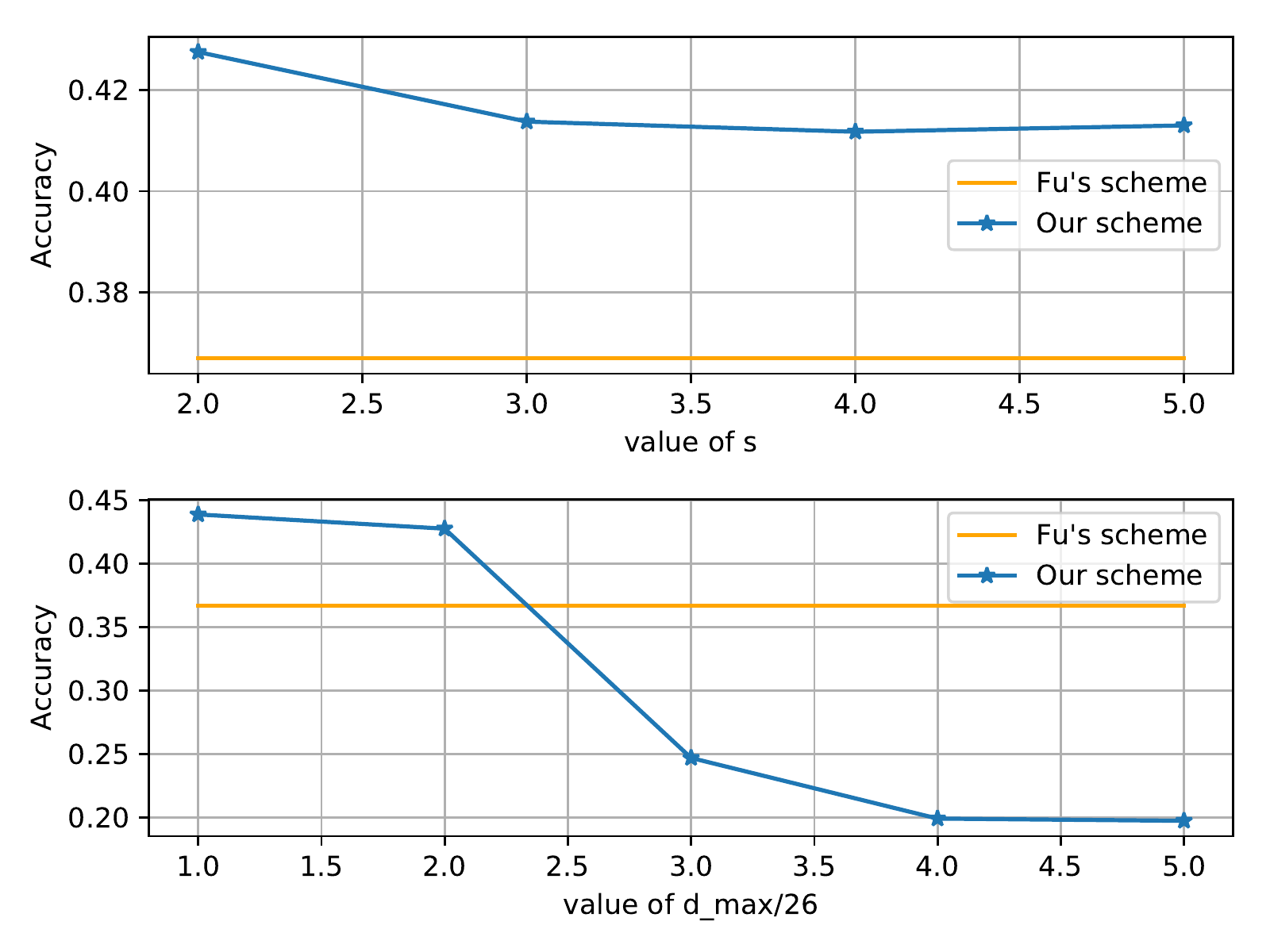}
	\captionof{figure}{Search accuracy under different value of $s$ and $u$ (= $d_{max} / 26$) for infection}
	\label{infectionexp}
\end{figure}

During the generation of OPU, \emph{``Infection''} is the most important mechanism to support fuzzy search for many different occasions as described in previous sections. It is one of the most important contributions in this paper. As the core of this step, ``Infection Function'' adopted is adjusted by two parameters: $u$ and $s$ in Eq \ref{infection}. We change the value of $u$ and $s$ independently to evaluate their influence on search performance. The experiment result is reported in Fig \ref{infectionexp}, where two misspelled keywords are inserted into each query string. The result argues that $s$ has a very slight influence on accuracy when varies from 2 to 5. However, when $u$ is set larger than 2, the accuracy reported in our scheme drops aggressively.

We provide an intuitive explanation for this phenomenon: a larger $u$ allows the weight sharing between two more distant positions in a keyword but in practice, data user misspells keywords by exchanging letters on distant positions in a much lower possibility. For example, it is more likely for the keyword ``listen'' to be misspelled to be ``lisetn'' than ``lestin''. Hence, if $u$ is set too larger, it may sacrifice the accuracy under most occasions to conform some rare extreme cases.

\begin{figure}[!htb]
	\centering
	\includegraphics[width=\linewidth]{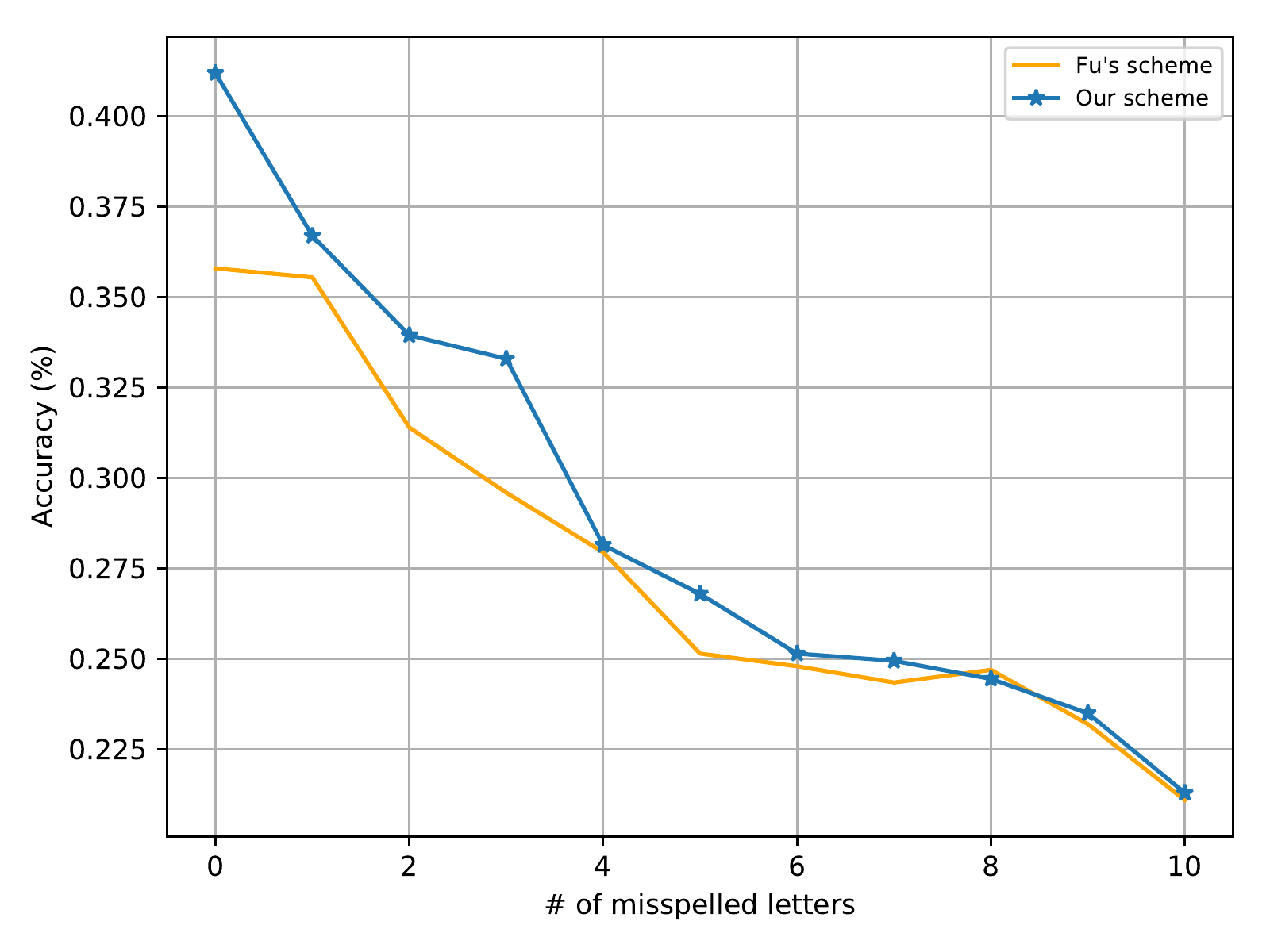}
	\captionof{figure}{Influence of misspellings on accuracy}
	\label{misspelling}
\end{figure}

To compare time efficiency in fuzzy search, we set two groups of experiments, one for spelling typos and the other for anagrams. For misspelling, each query consists of $Len_{query}$ keywords , and we select $m$ keywords in the query. The selected keywords become mutants by three types of letter operation on a single letter: $1)$ letter replacement $2)$ neighboring letters exchange and $3)$ deletion or addition a letter. For example, the mutants of the word ``search'' in three cases respectively could be ``seerch'', ``saerch'' and ``serch''/``searrch''. With the keyword mutant involved in queries, we imitate the real fuzzy search scenes in practice. The experiment result is reported in Figure \ref{misspelling}.

\begin{figure}[!htb]
	\centering
	\includegraphics[width=\linewidth]{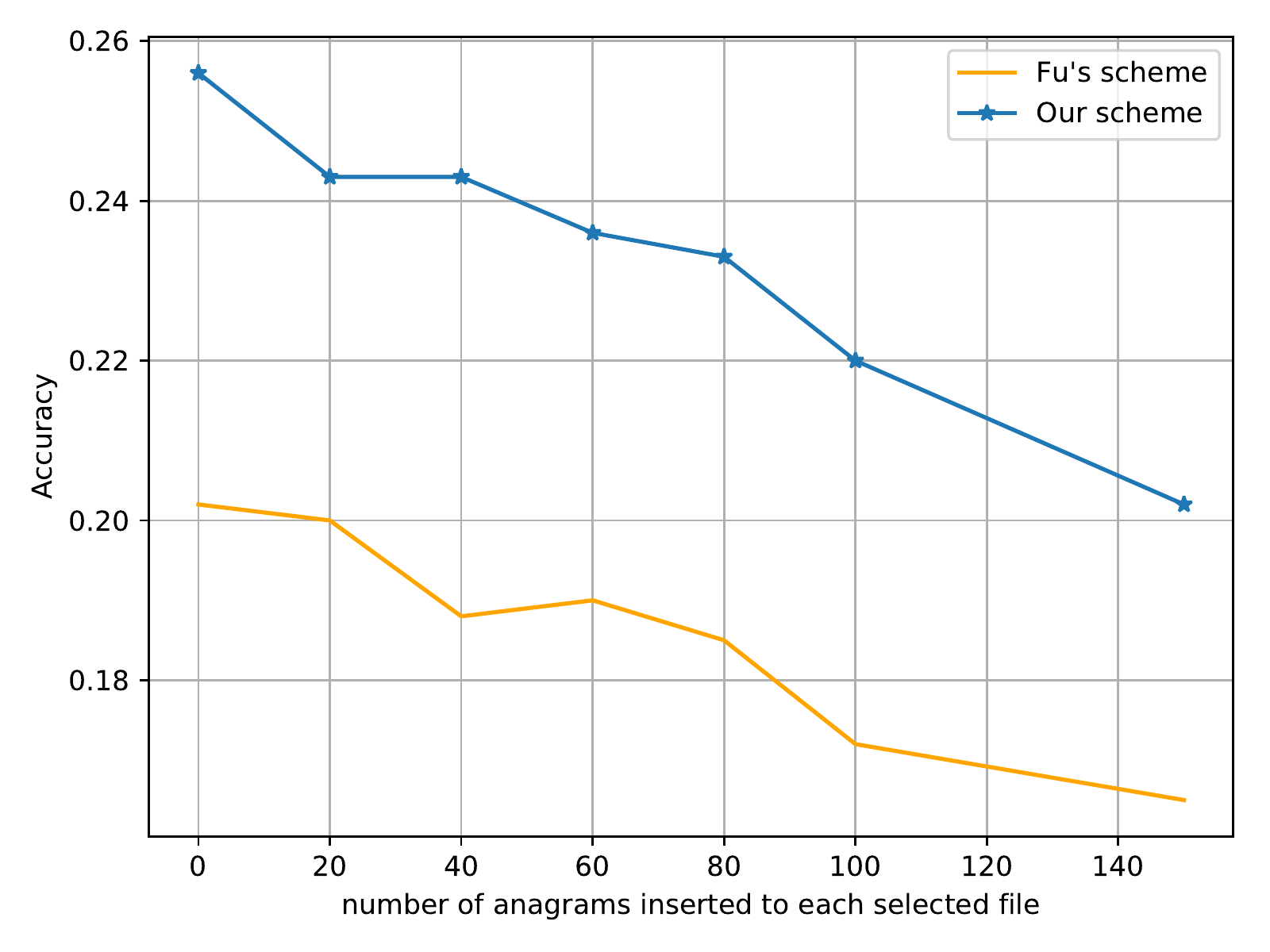}
	\captionof{figure}{Influence of number of file pairs to insert anagrams into ($F_A$) on accuracy}
	\label{anainsert}
\end{figure}

\begin{figure}[!htb]
	\centering
	\includegraphics[width=\linewidth]{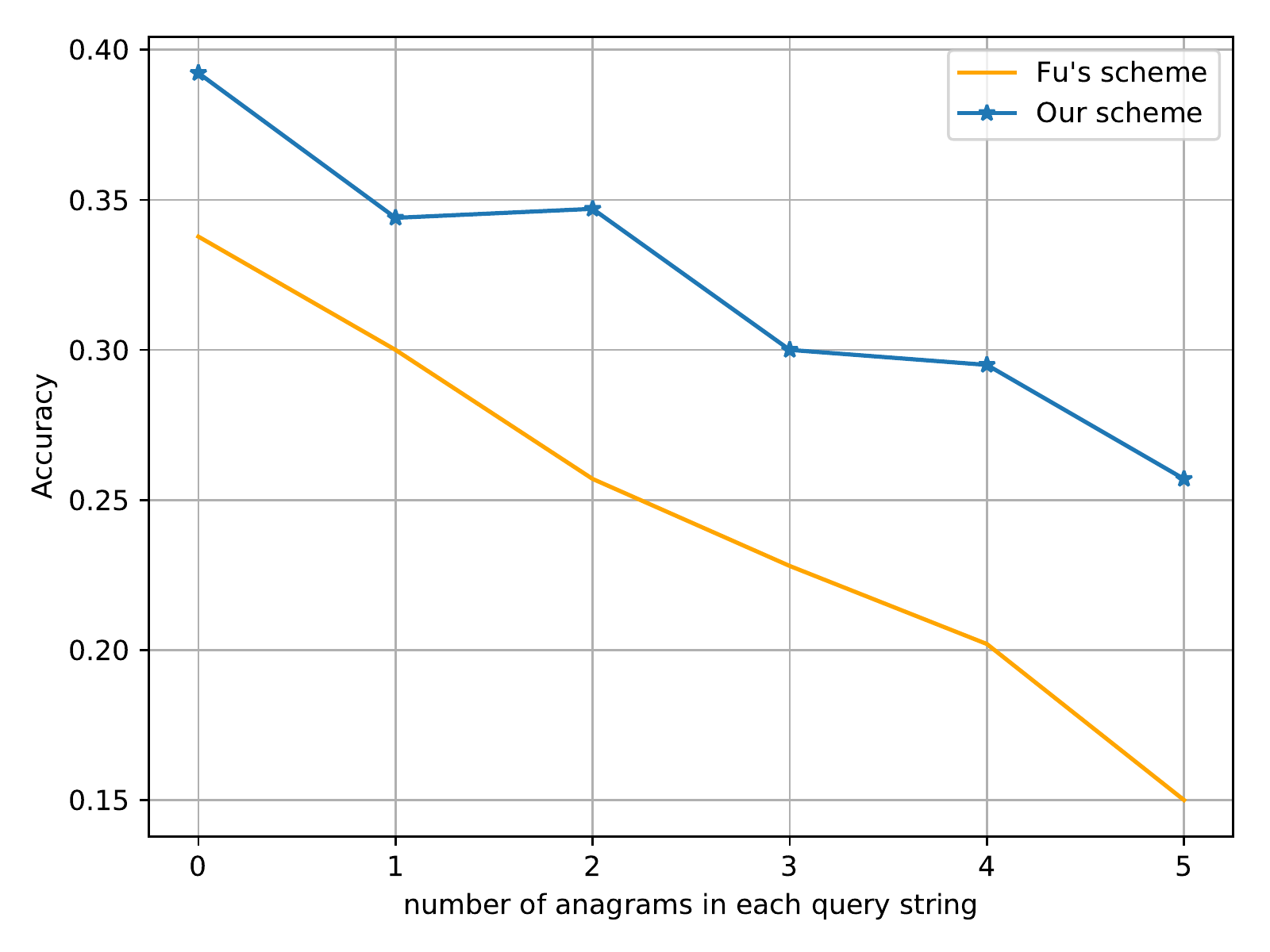}
	\captionof{figure}{Influence of number of anagrams in each query string ($N_A$) on accuracy}
	\label{anainquery}
\end{figure}

Another interesting case in fuzzy search is when anagrams are involved in queries and files. As explained already, anagrams make many currently popular search schemes on encrypted data invalid. To evaluate the scheme performance with anagrams involved, we collect in a total of 500 pairs of anagrams from \cite{anagramweb}, and then insert $N_A$ anagrams into files and queries in pair. Anagrams are inserted into $F_A$ files. $N_A$ and $F_A$ are adjusted separately in two groups of experiments for variable control. Results of experiments are reported respectively in Fig \ref{anainsert} and in Fig \ref{anainquery}.

Through these two groups of experiments, our scheme outperforms the scheme based on traditional 'uni-gram' more than for non-fuzzy search. We believe such improvement of accuracy obviously results from our proposed new mechanism to transform keywords in text to OPU.

\subsection{Time efficiency}

\begin{figure}[!htb]
	\centering
	\includegraphics[width=\linewidth]{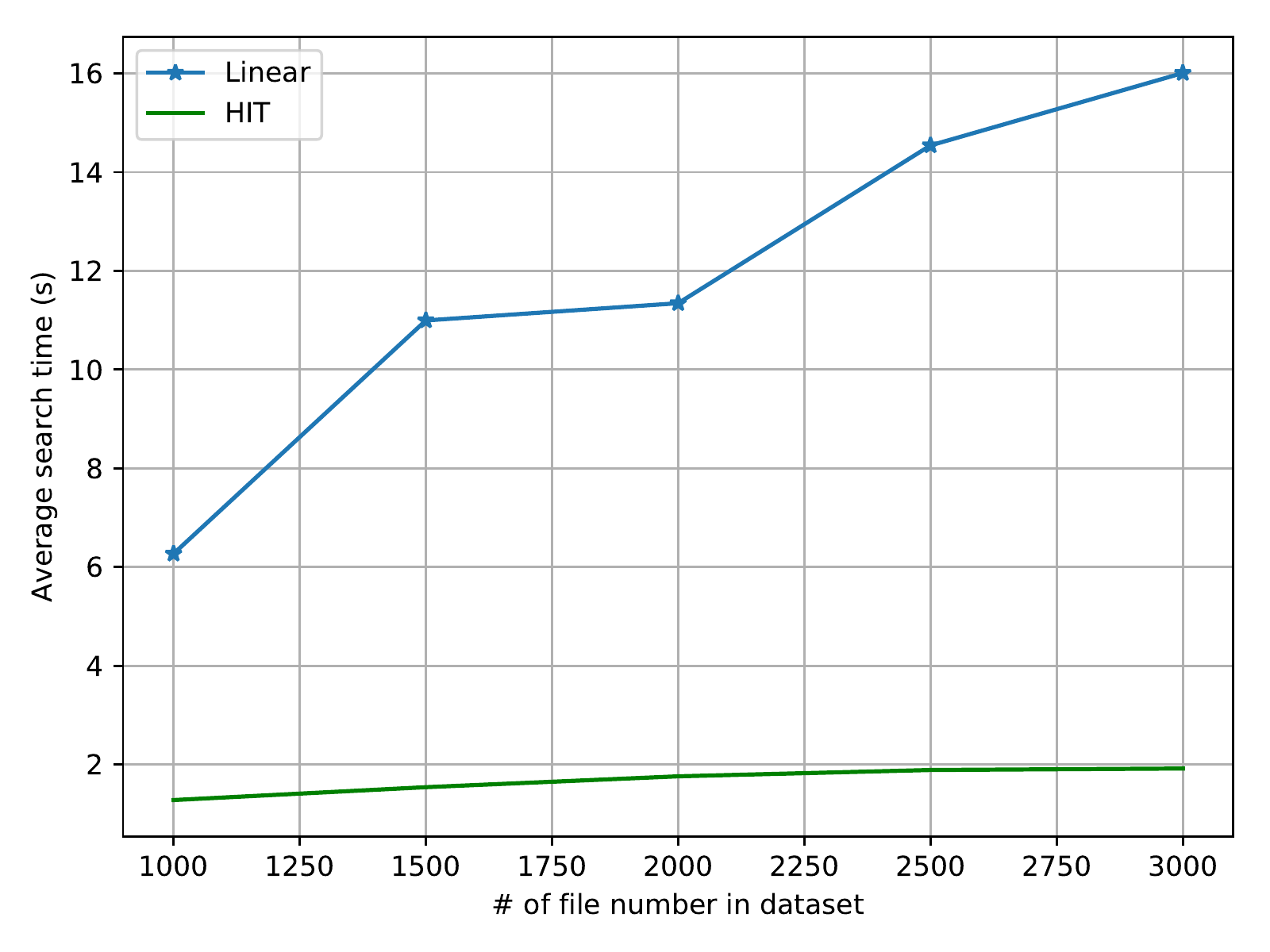}
	\captionof{figure}{Influence of the use of HIT  on search time}
	\label{searchtime}
\end{figure}

At last, we compare the search time consumption with HIT adopted and without it. From theoretical analysis, the time complexity is reduced from $\mathcal{O}{(n)}$ to $\mathcal{O}(log_n)$ and experiment result in Fig \ref{searchtime} conforms to the analysis well. Besides, comparing the result with previous experiments, we find HIT brings very slight harm to accuracy, which is acceptable in most practical occasions. 

% However, also not that the performance of HCI tree is highlt influence by the value of $e$ in Algorithm \ref{cluster}. Result in Fig. \ref{searchtime} is acquired with $e=0.4$.

\section{Conclusion}
In this paper, we propose a novel scheme for multi-keyword search over outsource encrypted data, where a fuzzy search is well supported in many different cases. Innovative proposal in the paper does no harm to the privacy guarantee on outsourced data. 

Our contributions can be concluded into two aspects: 
\begin{enumerate}
    \item we propose a novel keyword decomposing scheme based on what is named ``order-preserved uni-gram'' (OPU), which eliminates many weaknesses of previous ``uni-gram'' and ``n-gram'' schemes.
    \item we design a novel file indexing tree (HIT) which is based on hierarchical cluster tree while we improve the traditional K-means algorithm for data clustering. Thanks to the novel dynamic K-means algorithm, the indexing tree can be constructed more flexibleand with fewer parameters set manually.
\end{enumerate}

Experiments on real-world data prove the effectiveness of our proposed architecture. OPU brings accuracy improvement to a great extent under various experiment settings, and the proposed HIT increases search time efficiency without obvious hurt on search accuracy,

% or


%\appendices

%\section{}

%\section*{Acknowledgment}


\ifCLASSOPTIONcaptionsoff
  \newpage
\fi



\bibliographystyle{IEEEtran}
\bibliography{ref}
\end{document}